\documentclass[preprint,11pt]{revtex4-1}

\usepackage{amssymb,amsmath,setspace,graphicx,mathrsfs, float,slashed, mathrsfs, amsfonts, setspace, natbib, slashed}

\usepackage[usenames, dvipsnames]{color}

\begin{document}
	
\title{Sub-MeV Self Interacting Dark Matter}

\date{\today}

\author{Bhavesh Chauhan}
\email{bhavesh@prl.res.in}
\affiliation{Physical Research Laboratory, Ahmedabad}
\affiliation{Indian Institute of Technology, Gandhinagar, India}

\keywords{Light Dark Matter, Sterile Neutrino, Self Interacting}

\begin{abstract}
	In this paper, we present a model for sub-MeV dark matter with strong self interactions which can solve some of the small scale crisis of the $\Lambda$CDM. The dark matter is a Majorana fermion with only off-diagonal interactions with a hidden $U(1)_D$ gauge boson. The relic density is obtained by freeze-out of Boltzmann suppressed annihilations to a light fermionic species. The self interaction is a one loop process and constrained to be between 0.1 to 1 cm$^2$/g. Severe constraints from the BBN on $N_{eff}$ require that the dark and visible sector are not in thermal equilibrium during freeze-out. The effect of this temperature asymmetry is studied. 
\end{abstract}

\maketitle

\section{Introduction}


For the past few decades, we have extensively studied the gravitational interaction of Dark Matter (DM) and very little doubt remains of its existence (for an overview, see \cite{dmreview:garett,dmreview:bertone, dmreview:lisanti, dmreview:profumo, dmreview:phlen} and references therein). However, the particle nature of DM remains a mystery and we have no clue about its mass, spin, and interactions with other elementary particles. During the early days, Weakly Interacting Massive Particles (WIMPs) were postulated to be DM candidate but recent bounds from null results of terrestrial experiments have ruled out almost all of the interesting parameter space \cite{LUX2017}. Several new candidates have been proposed recently which get the correct relic abundance and are consistent with present detector bounds. \\

One of the simple solutions is to assume that the DM is light i.e. its mass is in the sub-GeV domain. In this limit, the local DM cannot produce sufficient recoil and thus will remain undetected in the traditional detectors. It has been proposed that electron recoil can be used to probe this parameter space \cite{subGeV:Essig,subGeV:Nanotubes,subGeV:3DDirac}. From the model building perspective, it was recently proposed that the 3-to-2 and 4-to-2 annihilations may be important for MeV and keV scale DM respectively \cite{simp:original}. Several interesting follow ups to this paradigm can be found in \cite{simp:1lest, simp:2bernal1, simp:4z3, simp:5nnlo, simp:6hambye, simp:7z2, simp:8murayama, simp:9ujjal, simp:10split, simp:11vector, simp:lhc}. One of the biggest issues with a sub-MeV DM is the conflict with the effective number of relativistic species ($N_{eff}$) during the Big-Bang-Nucleosynthesis (BBN) era \cite{rajendran}. To be consistent, one can assume that the dark sector has lower temperature than the Standard Model (SM) bath \cite{hidden:original, Foot1,Foot2,Foot3}, or that it freezes-in after the BBN \cite{berlin}.  \\

The standard model of cosmology, $\Lambda$CDM, has been hugely successful in explaining majority of the observed astrophysical phenomenon. However, the assumption of cold collision-less DM runs into what is dubbed as "small-scale crisis". The most prominent issues are the 'core vs. cusp' problem, the missing satellite problem, and "too-big-to-fail" problem. While individual resolutions to all the problems is possible, the assumption of self-interacting DM can solve some of these problems simultaneously \cite{sidm1,sidm2,sidm3,sidm4,sidm5,sidm6, sidm7, Khlopov1, Blinnikov1, Blinnikov2}. However, observation of galaxy cluster collisions puts a strong bound on this self interaction. For a recent review, one can refer to \cite{sidm:review} and references therein. For our analysis, we take the often used limit $\sigma_{SI}/m \sim 0.1 - 1 \text{ cm}^2/g$. \\

The outline of this paper is as follows. In Sec. \ref{model}, we define the low energy limit of the interaction Lagrangian and find the relic density and self-interaction in the model. In Sec. \ref{res}, we study the results and discuss the allowed parameter space before we conclude in Sec. \ref{conc}. 

\section{Model Description}
\label{model}

In this paper, we will consider the dark sector to be thermally decoupled from the Standard Model \cite{hidden:Das, BerlinPeV, SigurdsonHHDM, SIWDM}. The temperature asymmetry is characterised by the parameter $ \xi = (T_{d}/ T_{SM}) \leq 1$. Such a decoupling can be achieved if the interactions responsible for thermal equilibrium between the two sectors freeze out at high temperatures. In the absence of such interactions, one can postulate that the two sectors have been populated at different temperatures during reheating \cite{Asymmetric}. Because of this temperature asymmetry,  smaller mass for DM are allowed which is otherwise strictly constrained from the BBN $N_{eff}$. \\

We take DM to be Dirac fermion charged under a dark Abelian symmetry - $U(1)_D$. The gauge boson of this new symmetry, $Z^\prime$, acquires a mass from a high-scale spontaneous symmetry breaking. This transition is also responsible for generating a Majorana mass term which splits the dark fermion into two Majorana fermions ($\chi_1$ and $\chi_2$) with a mass gap  \cite{iDM1,iDM2,iDM3,iDM4,iDM5}. The lighter of the two Majorana states (say, $\chi_1$) will act as DM in this model.  In this mass basis, the coupling of $Z^\prime$ is purely off-diagonal as the Majorana states cannot carry any conserved quantum number. We add a light (almost massless) right-handed Dirac fermion ($f$)which is also charged under $U(1)_D$. The Majorana mass term for this light fermion can be avoided either by charge assignments or by assuming additional global symmetries. A detailed model is presented in the Appendix. \\

In the simplified picture, the interaction Lagrangian is given by, 
\begin{equation}
\label{lag} 
 \mathcal{L} \supset  -i g_D Z_\mu^\prime \left( \bar{\chi}_1 \gamma^\mu \chi_2 + \bar{f} \gamma^\mu f\right)
\end{equation}
where the coupling constant $g_D \approx 1$ ($\alpha_D = g_D^2/4\pi \approx 0.1$) for remainder of this paper. We assume the mass hierarchy
\begin{equation} 
 m_f \approx 0 \ll m_\chi=m_1 < m_2  = m_1(1 + \delta) \ll m_{Z^\prime}.
 \end{equation}
 As the fermions masses are in the sub-MeV domain and $\xi$ is not infinitesimally small , these particles contribute to the effective relativistic degrees of freedom during the BBN era as,  
\begin{equation}
N_{eff} = 3.046 + 2 \times \left(\frac{11}{4}\right)^{4/3} \xi^4. 
\end{equation}
The analysis of the Planck data indicated that $N_{eff} = 3.15 \pm 0.23$  \cite{Planck} which translates to $\xi \leq 0.45 (0.52)$ at $1\sigma (2 \sigma)$ level. However, if alternative cosmologies are taken into account, these constraints can be either severe or relaxed \cite{wCDM}. Hence, for our analysis, we take two bench mark scenarios $\xi = 0.5$ and $\xi = 0.3$ as we do not comment upon the source of this anisotropy. 

\subsection{Relic Density from Coannihilation} 

In this model, the relic density for $\chi_1$ is obtained from the coannihilations $\chi_1 \chi_2 \rightarrow \bar{f} f$. The importance of co-annihilations has been known for a long time \cite{Griest}, and novel applications were recently realised in \cite{BerlinGUT, Coann, OffDiag}. We follow the prescription in \cite{Griest} and important steps are mentioned for completeness. As $\chi_2$ can decay into $\chi_1$ via $\chi_2 \rightarrow \chi_1 \bar{f} f$, the coupled Boltzmann equations for tracking abundances of $\chi_1$ and $\chi_2$ are approximated by a single differential equation for the total number density $n = n_1 + n_2$ where $n_1$ and $n_2$ are the number densities of $\chi_1$ and $\chi_2$ respectively \cite{Edsjo}. During late times, $n$ is dominated by $n_1$ as most of $\chi_2$ has decayed. The Boltzmann equation for $n$ is, 
\begin{equation}
\label{boltz:n}
\dfrac{dn}{dt} + 3 H n 	= - \langle \sigma v \rangle_{eff} (n^2 - \bar{n}^2).
\end{equation}
where bar indicates the equilibrium density and, 
\begin{equation}
\langle \sigma v \rangle_{eff}   = \sum_{i j} \langle \sigma_{ij} v_{ij} \rangle \frac{\bar{n}_i \bar{n}_j}{\bar{n}^2}. 
\end{equation}

\begin{figure}
	\includegraphics[width=6 cm]{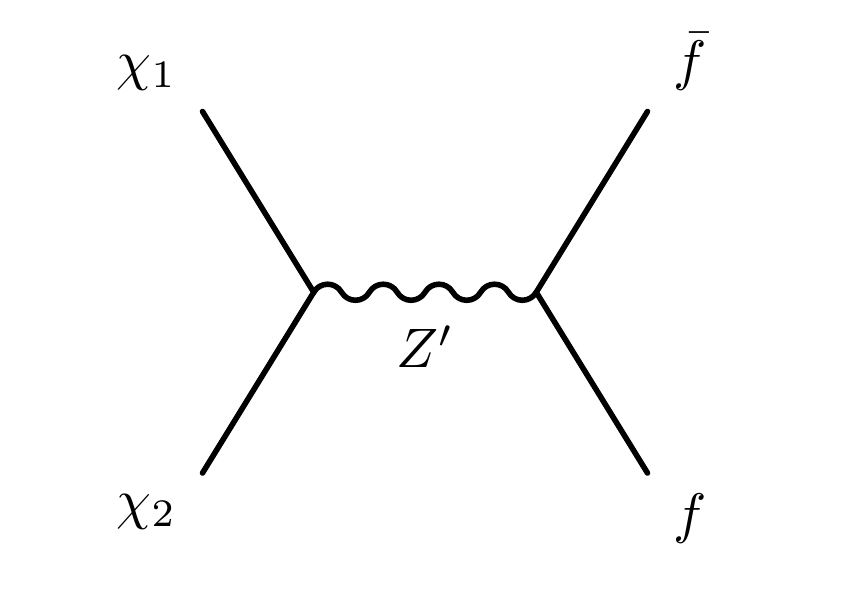}
	\caption{\label{fig:relic} The annihilation channel for $\chi_1$ whose freeze-out determines the relic density}
\end{figure}
Due to the off-diagonal interactions of $Z^\prime$, processes such as $\chi_1 \chi_1 \rightarrow \bar{f} f $ are forbidden at tree level, and the only annihilation channel is  $\chi_1 \chi_2 \rightarrow \bar{f} f $. Thus the effective cross section is given as, 
\begin{equation}
\langle \sigma v \rangle_{eff}   = 2  \langle \sigma_{12} v_{12} \rangle \frac{\bar{n}_1 \bar{n}_2}{\bar{n}^2} \approx 2  \langle \sigma_{12} v_{12} \rangle \frac{\bar{n}_2 }{\bar{n}_1}
\end{equation}
where the approximation obtained by using $\bar{n}_2 \ll \bar{n}_1$ is only indicative and we use full expression for the numerical analysis. Recently, utilization of such Boltzmann suppression for Light DM has been realised in \cite{light:1:forbidden, light:2:fourth} however with a small mass-gap ($\delta<1$). In this paper, we have considered a significantly large mass gap between the two states ($\delta \sim 2-6$). We use the following expression for number density, 
\begin{equation}
n_i (m, T) = \frac{T}{2 \pi^2} m^2 K_2\left(\frac{m}{T}\right)
\end{equation}
and the thermal averaged cross section in the s-wave limit is given as, 
\begin{equation}
\langle \sigma_{12} v_{12} \rangle = \frac{1}{32 \pi} \frac{g_D^4}{m_{Z^\prime}^4} \left(m_1 + m_2\right)^2
\end{equation}
One can rewrite \eqref{boltz:n} using the abundance $Y = n/s$ where $s$ denotes the total entropy density of standard model and the dark sector. As $\xi<1$, the entropy is dominated by the SM bath and to a very good approximation, 
$$s \approx s_{SM} = \frac{2 \pi^2}{45} g_\ast^s(T_{SM}) T_{SM}^3.$$
The equilibrium abundance is given by,
\begin{equation}
\label{yeq}
\bar{Y}(x, \xi) = \xi^3 \frac{d_\chi}{g_\ast^s(m_\chi/x \xi)} \frac{45}{4 \pi^4} x^2 K_2(x)
\end{equation} 
where $x = m_\chi /T_d$ is a measure of the dark sector temperature. The freeze out occurs when, 
\begin{equation}
\label{condition}
\left[ \bar{n} \langle \sigma v \rangle_{eff}\right] _{x_f} =  H(\xi, x_f)
\end{equation}
i.e. when the interaction rate becomes less than the Hubble Rate $H = 1.66 g_\ast(T) T^2/M_{pl} = 1.66 g_\ast(T/\xi) m_\chi^2/ (x^2 \xi ^2 M_{pl})$. The present day abundance, $Y_\infty$, is given as, 
\begin{equation}
\label{yinf}
Y_\infty = \frac{  c \bar{Y}(x_f, \xi) }{ 1 + \lambda J(x_f) c \bar{Y}(x_f, \xi)}
\end{equation}
where $c, \lambda,$ and $J(x_f)$ are defined in Appendix A. The relic density of DM is given by, 
\begin{equation}
\label{relic}
\Omega h^2 = m_\chi s_0 Y_\infty \frac{h^2}{\rho_c} \approx 282  \left( \frac{m_\chi}{\text{keV}} \right) \left( \frac{T_\gamma}{2.75~K}\right)^3 c~\bar{Y}(x_f, \xi)
\end{equation}
where the approximation is true in the limit $ \lambda J(x_f) c \bar{Y}(x_f, \xi) \ll 1$. We use \eqref{condition} to numerically determine the freeze-out temperature and enforce that $x_f \geq 3$ so that the non-relativistic approximation is valid. This restricts us from taking smaller values for $\xi$ and $m_1$. Then we determine the relic density using \eqref{yeq}, \eqref{yinf}, \eqref{relic}, and compare with the observed value from Planck \cite{Planck}, 
\begin{equation}
\Omega_\chi h^2 = 0.118 \pm 0.002.
\end{equation}
Understanding that such an estimate is only an approximation to solving the complete Boltzmann equations, we conservatively take an error on 5\% in our analysis.

\subsection{One-Loop self interaction }

\begin{figure}[t]
	\centering
	\includegraphics[width=7cm]{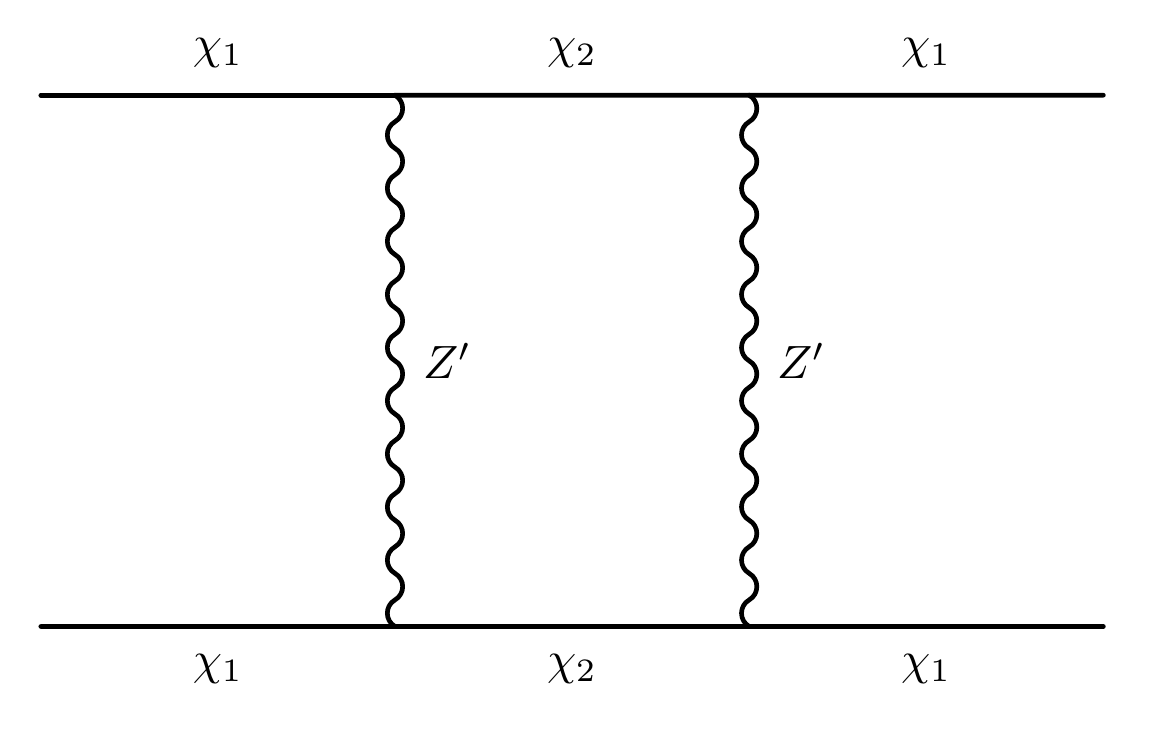}
	\caption{\label{fig:SI} The Feynman diagram for the self interaction of DM. There are seven other "crossed" diagrams.}
\end{figure}

\begin{figure}[H]
 	\centering
 	\includegraphics[width=14cm]{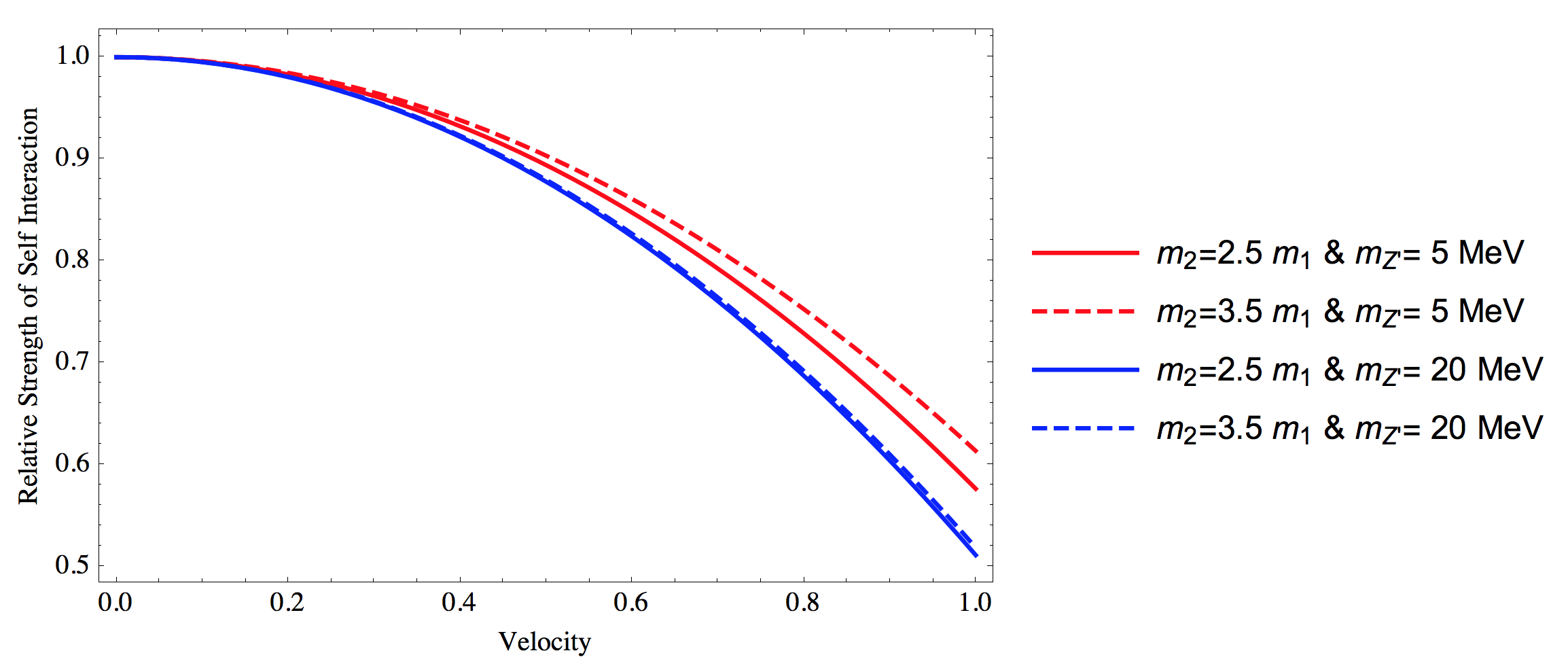}
 	\caption{\label{fig:RelStr} The relative strength of self interactions i.e. $\sigma_{SI}(v) / \sigma_{SI}(0)$ is shown as a function of velocity for various choices of parameters. }
 \end{figure}

One of the features of this model is that the self-interaction of Dark Matter is not a tree level process. At one loop level, there are eight diagrams that contribute to $\chi_1 \chi_1 \rightarrow \chi_1 \chi_1$ when $\chi_2~\text{and}~Z^\prime$ are in the loop. A representative diagram is shown in Fig. \ref{fig:SI}. In \cite{SSIDM, SSIDM2},  the self interaction of inelastic DM was studied in the limit of large $m_\chi$ and light propagator. In this study, we calculate the self interaction in the limit of small $m_\chi$ and heavy propagator. Since the loop particles are significantly heavier than the external ones, we use the decoupling limit where we ignore the external momenta while evaluating the loop. We use Package-X \cite{PackageX} and the Unitary Gauge to calculate the loop function and the cross section. It was checked that the infinities cancel systematically and we are left with a finite part. The self-interaction cross section in the s-wave approximation is given as, 
\begin{equation}
\frac{\sigma_{SI}}{m_1} = \frac{9}{256 \pi^5 }g_D{}^8 \frac{  m_1 \left(m_2{}^6+3  m_2{}^2m_{Z'}{}^4+6
	m_2{}^2m_{Z'}{}^4 \log \left(\frac{m_2{}^2}{m_{Z'}{}^2}\right) -4
	m_{Z'}{}^6\right){}^2}{m_{Z'}{}^4 \left(m_{Z'}{}^2 - m_2{}^2\right){}^6}.
\end{equation}
The calculation is detailed in Appendix B. The velocity dependence of the self interaction is shown in Fig. \ref{fig:RelStr}. It can be seen that the change is very small for non-relativistic case ($v < 0.1 c$). Therefore, we use the estimate $\frac{\sigma_{SI}(0)}{m_1} = 0.1 -1~cm^2/g$ to constrain the parameter space. 

\subsection{A comment on the light fermion}

One of the crucial assumption of this model is the existence of a massless fermionic species. One of the possibilities is that it is a part of the radiation component today albeit, the strong self interactions would prevent it from being Hot Dark Matter candidate. The other interesting possibility is that it is a sterile neutrino which also mixes with the active neutrinos. It has been pointed out that in presence of self-interactions, the sterile neutrino acquires a large thermal mass in the early universe and the mixing is suppressed \cite{sterile:Dasgupta, sterile:Hannestad}. This allows one to have larger mixing angles in the present era and helps resolving some of the short-baseline neutrino anomalies \cite{LSND}. However, to avoid DM-neutrino scattering in the early universe, we require much smaller vacuum mixing angles that cannot explain these anomalies, but can be probed in future experiments. \\

The role of the light fermion in cosmology would be similar to that of dark radiation. The most stringent bounds on dark radiation comes from BBN $N_{eff}$ which we have considered already. As this light fermion is part of a secluded and colder sector, it plays very little role in structure formation.  \\

\section{Results and Discussion}
\label{res}
As pointed out before, we take $g_D \approx 1$ for our analysis. This is a domain where the interactions are strong but perturbativity still holds. In \cite{Yeche}, bounds on mass of Warm Dark Matter from Lyman-$\alpha$ is determined to be $M_{WDM} \geq \text{few keV}$. We only consider $m_1 > 10$ keV in this work. We analyse the parameter space of $\delta - m_{Z^\prime}$ for various masses of $m_1 \in \{ 10\text{ keV}, 1 \text{ MeV} \}$ that give the correct relic density and self-interactions. As the self-interactions do not depend on $\xi$, one can see that the limits are same for the two benchmark cases. It is to be noted that a heavier $Z^\prime$ is associated with smaller self interaction. \\

 The dependence of relic density on $\xi$ can be simply understood as follows. From \eqref{yeq} one can see that $\bar{Y}$ is a monotonically decreasing function of $x$. To compensate for small $\xi$, one needs a smaller $x_f$. This means that the effective cross section should be smaller such that freeze-out occurs earlier. This smallness is brought by a larger Boltzmann suppression due to heavier $m_2$. In an analogous way, one can argue the dependence of the relic density on $c$. \\
 
 As the DM is part of a secluded sector, one does not anticipate any signals in Direct Detection experiments and colliders. This is consistent with the present status of these terrestrial experiments. Such a dark matter can only have gravitational signatures and can be probed through structure formation. Due to the self interactions, the DM behaves as WDM and is consistent with the present understanding. In future, as the limits on BBN $N_{eff}$ are tightened, there will be less parameter space for the model to thrive.  

\begin{figure}
	\includegraphics[width= 14cm]{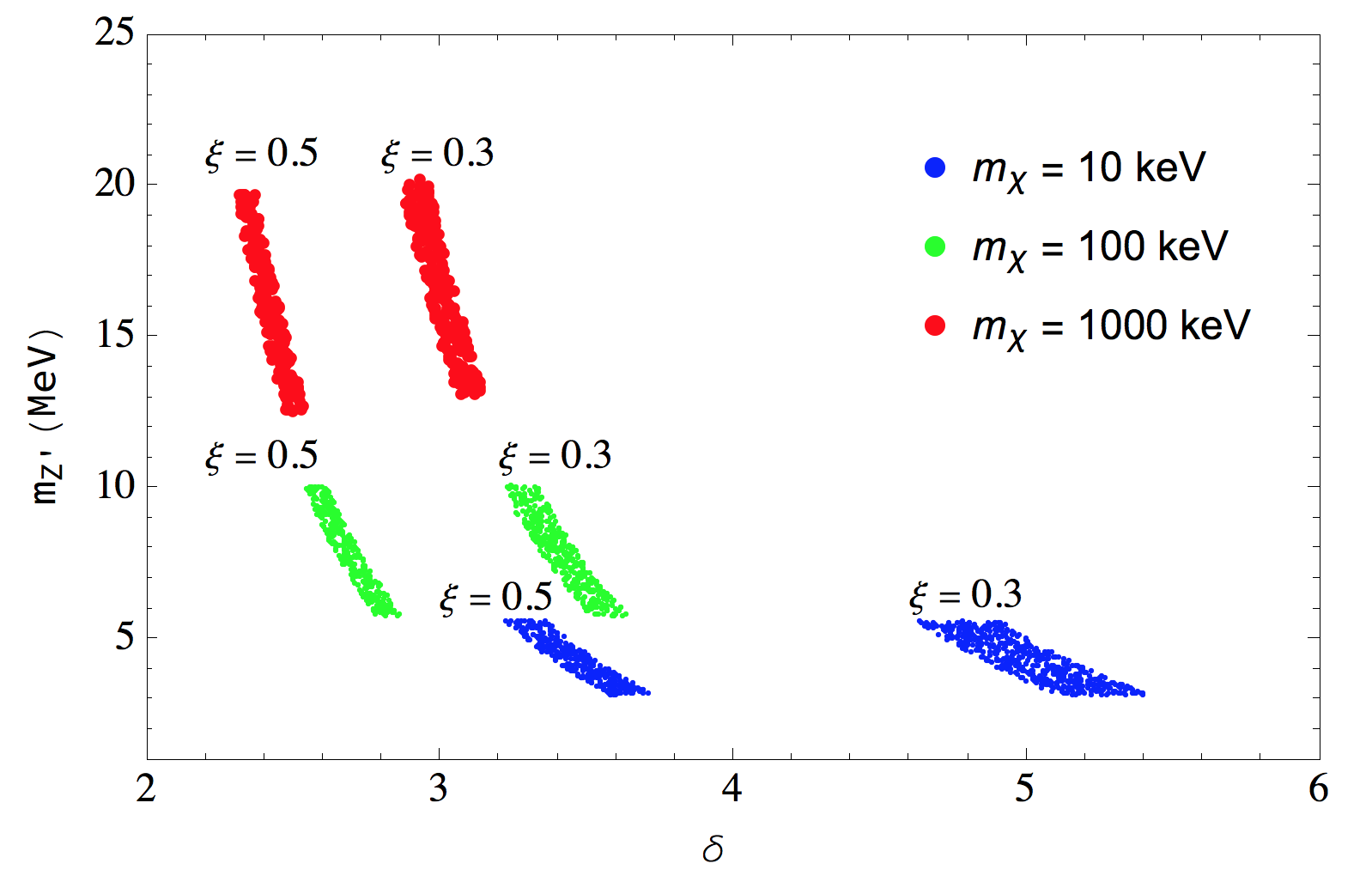}
	\caption{\label{fig:result} The allowed parameter space for $m_\chi= $ 10 keV (Blue), 100 keV (Green), and 1 MeV (Red) is shown for benchmark models $\xi = 0.5$ (left) and $\xi = 0.3$ (right). The upper (lower) limit of $m_{Z^\prime}$ corresponds to $\sigma_{SI}/m_1 = 0.1 (1.) \text{ cm}^2/g$.}
\end{figure}

\newpage
\section{Conclusion}
\label{conc}
In this paper, we have seen that one can get the correct relic density and appropriate self-interactions for a sub-MeV Dark Matter if it has strong off-diagonal interactions with a heavier spin-1 boson. The annihilation cross section is Boltzmann-suppressed and the self interaction is loop-suppressed thus allowing the mass scales to go as low as $\mathcal{O}(10)$ keV while keeping the gauge coupling constant naturally large. Such a light DM must be part of a decoupled sector at a lower temperature in order to be consistent with BBN.

\subsection*{Appendix A: Calculation of Relic Abundance}
The calculation of relic abundance of dark matter has been excellently treated in the book "The Early Universe" by E. W. Kolb and M. S. Turner \cite{KolbTurner}. We follow the general prescription laid by them while making necessary changes due to the temperature asymmetry. Similar calculation is performed in \cite{hidden:original} and the only difference is that we use hidden sector temperature to define $x$ while using SM entropy to define abundance $Y$. Such a definition is advantageous in models where the hidden sector entropy is not conserved explicitly (e.g. when a minor component decays into SM particles during late times). In such scenarios, the total entropy density, which is mainly SM entropy, is a good proxy for dilution effect. Otherwise, the treatment is analogous and one can use either definitions. \\

The Boltzmann equation for the total number density \eqref{boltz:n} can conveniently expressed in terms of the abundance 
\begin{equation}
Y = \frac{n}{s}
\end{equation}
which is free from the dilution due to expansion. Note that $s$ denotes the total entropy density of the dark and visible sectors. However, due to temperature asymmetry, one can ignore the contribution from the dark sector. Also note that, since the total entropy is conserved, $\dot{s} + 3 H s = 0$. During the radiation dominated era, the scale factor $R \sim t^{1/2}$ which gives us,
\begin{equation}
\dfrac{dx}{dt} = \frac{\tilde{H}(m_\chi, \xi) }{x}
\end{equation}
where $x = m_\chi/ T_d$ and in terms of Planck Mass $M_{pl} = 1.22 \times 10^{25}$ keV
\begin{equation}
\tilde{H}(m_\chi, \xi) = 1.66 \sqrt{ g_\star \left(\frac{m_\chi}{x \xi}\right)} \frac{1}{\xi^2} \frac{m_\chi^2}{M_{pl}}. 
\end{equation}
Using \eqref{yeq} and, 
\begin{equation}
\tilde{s}(m_\chi, \xi) = \frac{2 \pi^2}{45} g_\star^s \left(\frac{m_\chi}{x \xi}\right) \frac{m^3}{\xi^3}, 
\end{equation}
the Boltzmann equation for abundance is, 
\begin{equation}
\label{boltz:y}
\dfrac{dY}{dx} = - \frac{ \tilde{s}}{\tilde{H}} \frac{\langle \sigma v \rangle_{eff}}{x^2} \left( Y^2 - \bar{Y}^2\right).
 \end{equation}
 Note that the temperature (hence, $x$) dependence in the effective cross section comes only from the Boltzmann factor and hence one can write, 
 \begin{equation}
 \langle \sigma v \rangle_{eff} = \sigma_0 f(x, \delta)
 \end{equation}
where, 
\begin{equation}
f(x, \delta) =\frac{ (1 + \delta)^2 K_2(x) K_2( (1 + \delta) x)}{(K_2(x) +  (1 + \delta)^2 K_2( (1 + \delta) x))^2}. 
\end{equation} 
 Using the dimensionless quantity $\lambda = \sigma_0 \tilde{s} / \tilde{H}$ one can simplify \eqref{boltz:y} as, 
 \begin{equation}
 \label{boltz:y2}
 \dfrac{dY}{dx} = - \lambda \frac{f(x, \delta)}{x^2} \left( Y^2 - \bar{Y}^2\right).
 \end{equation}
 which can be further simplified using the difference $\Delta = Y - \bar{Y}$ and approximately solved when $x \gg x_f$ and $\Delta \approx Y \gg \bar{Y}$ which gives, 
 	\begin{equation}
 	\label{boltz:delta}
 	\Delta^\prime \approxeq - \lambda \frac{f(x,\delta)}{x^2} \Delta^2.
 	\end{equation}
 Upon integration from freeze-out to the present day of \eqref{boltz:delta}, we get,
 \begin{equation}
 \frac{1}{Y_\infty} = \frac{1}{\Delta_\infty} = \frac{1}{\Delta_f} + \lambda \int_{x_f}^{\infty} \frac{f(x, \delta)}{x^2} dx  =  \frac{1}{\Delta_f} + \lambda J
 \end{equation} 
and the $J$ integral can be performed numerically once $x_f$ is determined. It was shown in \cite{hidden:original}, that the approximation 
\begin{equation}
\Delta_f = c \bar{Y}(x_f, \xi)
\end{equation}
 agrees with the numerical solution of \eqref{boltz:y2} if c = 0.2 (0.5) for $\xi = 0.3 (0.8)$. This gives us the final result, 
 \begin{equation}
 \label{app:yinf}
 Y_\infty = \frac{  c \bar{Y}(x_f, \xi) }{ 1 + \lambda J(x_f) c \bar{Y}(x_f, \xi)}
 \end{equation}
 which is was shown in \eqref{yinf}. For our analysis, we assume $c = 0.2$ and note that any change in $c$ will proportionately scale the relic density. 
 
\subsection*{Appendix B: Calculation of Self Interaction}

We calculate the amplitude for the process, 
\begin{equation}
\chi_1 (p_1)  + \chi_2 (p_2) \rightarrow \chi_1 (k_1)  + \chi_2 (k_2) 
\end{equation}
where $p_i$ and $k_i$ are the four momentum of the particles. There are eight Feynman diagrams for this process which are related by crossing to the one shown in Fig. \ref{fig:SI}. In the decoupling limit, the amplitude is 
\begin{equation}
\mathcal{M}_1 \sim \frac{g^4 \left[ \bar{u}(k_1) \gamma^\mu (\slashed{q} + m_2) \gamma^\alpha u(p_1)\right] \left[ \bar{v}(p_2) \gamma^\beta (\slashed{q} + m_2) \gamma^\nu v(k_2)\right]P_{\alpha \beta} P_{\mu \nu}}{(q^2 - m_{Z^\prime}^2) (q^2 - m_{2}^2)^2}
\end{equation}
where $q$ is the loop momentum and $P_{\mu \nu}$  in the Unitary gauge is given by, 
\begin{equation}
P_{\mu \nu} =  - g_{\mu \nu} + \frac{q_ \mu q_\nu}{m_{Z^\prime}^2}.
\end{equation}
The other crossed amplitudes ($\mathcal{M}_2 \rightarrow \mathcal{M}_6$) are related to $\mathcal{M}_1$ by $\beta  \leftrightarrow \mu, \beta \leftrightarrow \nu, k_1 \leftrightarrow k_2$. There are two diagrams pertaining to the colloquial "s-channel" due to Majorana nature of the incoming fermions. The relative sign of graphs must be taken correctly for cancellation of the infinities. One can evaluate the loop-integral using Package-X or any other alternative. The final result can be simply expressed in the $\{S, V, T, A, P \}$ basis as, 
\begin{equation}
\mathcal{M} =g^4  \sum_{i = S,V,T,A,P}^{} \left( C_i \left[ \bar{u}(k_1) \Gamma_i u(p_1)\right] \left[ \bar{v}(p_2) \Gamma_i v(k_2)\right] +  C^\prime_i \left[ \bar{v}(p_2) \Gamma_i u(p_1)\right] \left[ \bar{u}(k_2) \Gamma_i v(k_1)\right] \right)
\end{equation}
Note that the mixed terms (e.g $V-A$) are absent. The only non-zero coefficients are 
\begin{equation}
C_A = \frac{6 m_2{}^2 m_{Z'}{}^2 \log
	\left(\frac{m_2{}^2}{m_{Z'}{}^2}\right)}{\left(m_2{}^2-m_{Z'}{}^2\right){}^3}-\frac{3 \left(m_2{}^4-m_2{}^2 m_{Z'}{}^2+2 m_{Z'}{}^4\right)}{m_{Z'}{}^2
	\left(m_2{}^2-m_{Z'}{}^2\right){}^2}
\end{equation}

\begin{equation}
C_T = -\frac{m_2{}^2 \left(m_2{}^2-3 m_{Z'}{}^2\right)}{m_{Z'}{}^2
	\left(m_2{}^2-m_{Z'}{}^2\right){}^2}-\frac{2 m_2{}^2 m_{Z'}{}^2 \log
	\left(\frac{m_2{}^2}{m_{Z'}{}^2}\right)}{\left(m_2{}^2-m_{Z'}{}^2\right){}^3}
\end{equation}

\begin{equation}
C^\prime_S = \frac{6 m_2{}^2 \left(m_2{}^2-3 m_{Z'}{}^2\right)}{m_{Z'}{}^2
	\left(m_2{}^2-m_{Z'}{}^2\right){}^2}+\frac{12 m_2{}^2 m_{Z'}{}^2 \log
	\left(\frac{m_2{}^2}{m_{Z'}{}^2}\right)}{\left(m_2{}^2-m_{Z'}{}^2\right){}^3}
\end{equation}

\begin{equation}
C^\prime_A = \frac{3 m_2{}^2 m_{Z'}{}^2 \log
	\left(\frac{m_2{}^2}{m_{Z'}{}^2}\right)}{\left(m_2{}^2-m_{Z'}{}^2\right){}^3}-\frac{3 \left(m_2{}^4-m_2{}^2 m_{Z'}{}^2+2 m_{Z'}{}^4\right)}{2 m_{Z'}{}^2
	\left(m_2{}^2-m_{Z'}{}^2\right){}^2}
\end{equation}
In terms of these coefficients, the non-relativistic squared amplitude is
\begin{equation}
\overline{|\mathcal{M}|^2} = 16 m_1^4 \left( 3 C_A + 2 C^\prime_A - 6 C_T \right)^2  - 16 m_1^4 v^2\left(C_A + 2 C^\prime_A - 6 C_T \right) \left(3 C_A + 2 C^\prime_A - 6 C_T \right) 
\end{equation}
and the transfer cross section for self interaction is
\begin{equation}
\sigma_{SI} = \int d\Omega (1 - \cos(\theta)) \left( \frac{d \sigma}{d \Omega} = \frac{1}{64 \pi^2 (4 m_\chi^2)} \overline{|\mathcal{M}|^2}\right) \approx \frac{1}{64 \pi m_\chi^2}\overline{|\mathcal{M}|^2}
\end{equation}

\subsection*{Appendix C: Possible UV Completion}

In this section we consider a possible UV completion of the simplified model presented above. The standard model gauge group is extended by an $U(1)_D$ symmetry. We add four fermions and a scalar to the model which are singlets under SM gauge symmetry. Their charges under the new symmetry are given in Table I. 

\begin{table}[H]
	\centering
	\begin{tabular}{|c | c | c | c | c | c |}
		\hline
		Fields & $\psi_1$ & $\psi_2$ & $f_1$ & $f_2$ & $\phi$ \\
		\hline
		$Q_D$ & 1 & -1 & $a$ & $-a$ & 2 \\
		\hline
	\end{tabular}
	\caption{The new fields in the dark sector and their charges under $U(1)_D$ symmetry.}
\end{table}

The above choice of charges assures that the model is anomaly free. One can chose $a \approx1$ but $\neq 1$ to ensure that $\phi$ does not have Yukawa like interaction with $f_1$ or $f_2$. The most general Lagrangian for the dark sector is, 
\begin{align}
 \mathcal{L} &= \bar{\psi}_1 ( \slashed{D} - m) \psi_1 + \bar{\psi}_2 ( \slashed{D} - m) \psi_2 + \bar{f}_1 ( \slashed{D} - m_f) f_1 + \bar{f}_2 ( \slashed{D} - M_f) f_2  \\
 &+ y \phi \bar{\psi}_1 \psi_2 + h.c. \\
 &+ (D_\mu \phi)^\dagger (D^\mu \phi)  - \frac{1}{4} X^{\mu \nu}X_{\mu \nu} \\
 &+ \frac{\epsilon}{4}X^{\mu \nu}F_{\mu \nu} + \eta \phi^\dagger \phi H^\dagger H \\
 &- \mathcal{V}(\phi)
\end{align}
where $H$ is the SM Higgs' field, $X_{\mu \nu} = \partial_\mu Z^\prime_\nu - \partial_\nu Z^\prime_\mu$ is the field strength for the $Z^\prime$, and 
\begin{equation}
D_\mu = \partial_\mu - i g_D Q_D Z^\prime_\mu
\end{equation}
is the gauge covariant derivative. To begin with, we consider the limit where $\epsilon \rightarrow 0,$ and $\eta \rightarrow 0$ which is motivated from the assumption that the dark sector is thermally secluded from the visible sector. Also, these interactions cannot be generated via loops which allows us to take their coefficients to be vanishingly small. \\

The potential for the new scalar field has the usual form considered for spontaneous symmetry breaking. 
\begin{equation}
\mathcal{V} (\phi) = - \mu^2 \phi^\dagger \phi +  \lambda ( \phi^\dagger \phi)^2
\end{equation}
The symmetry breaking not only gives mass to the new gauge boson, but also generates an off diagonal mass term from the Yukawa-like interaction.  In the $\psi_1 - \psi_2$ basis, the mass matrix is, 
\begin{equation}
\hat{M} = \begin{pmatrix}
m & y v_\phi \\
y v_\phi & m \\
\end{pmatrix}
\end{equation}
which has eigenvalues $m \pm y v_\phi$.  One can go to the mass eigenstates by the transformation, 
\begin{equation}
\psi_1 \rightarrow \frac{\chi_1 + \chi_2}{\sqrt{2}} ~ and~\psi_2 \rightarrow \frac{\chi_1 - \chi_2}{\sqrt{2}}.
\end{equation}
The Lagrangian for $\chi_1$ and $ \chi_2$ is, 
\begin{align}
\mathcal{L} &= \bar{\chi}_1 ( \slashed{\partial} - m_1) \chi_1 +\bar{\chi_2} ( \slashed{\partial} - m_2) \chi_2 +  i g_D (  \bar{\chi}_1  \slashed{Z}^\prime \chi_2 + \bar{\chi}_2  \slashed{Z}^\prime \chi_1 ) + ...
\end{align}
where the ellipses denote interactions with the Higgs' scalar of the dark sector. In terms of the free parameters, one can fix $v_\phi$ given the mass of the $Z^\prime$ boson. However, by varying $\lambda$ one can make the scalar sufficiently heavy such that it does not affect the low scale dynamics. Also, one can speculate that if there are other heavy fields in the dark sector, there may be large radiative corrections to the scalar mass. The mass gap between the two states is determined by the Yukawa coupling ($m_2 = m_1 + 2 y v_\phi$) and can be considered as a free parameter. In the limit $M_f >> m_1,m_2$, this model essentially reduced to the one considered in the paper.  

\newpage
\noindent \textbf{Acknowledgements:} 
The author would like to thank Prof. Subhendra Mohanty and Prof. Namit Mahajan for several useful discussions and their insights. The author is also grateful to Arnab Dasgupta for going through the manuscript and providing valuable suggestions. The author also thanks the anonymous referee for pointing out important constraints.

\end{document}